\documentstyle[aps,epsf,twocolumn]{revtex}

\begin{document}
\twocolumn[
\begin{center}
\vspace{-5mm}
{\large{\bf Quasiparticles of ${\mbox{\boldmath $d$}}$-wave superconductors
in finite magnetic fields}}

\vspace{5mm}
{Kouji Yasui
and Takafumi Kita}
\par

{\small\em Division of Physics, Hokkaido University, 
Sapporo 060-0810, Japan}
\par

({\today})
\vspace{3mm}

{\small \parbox{142mm}{\hspace*{5mm}
We study quasiparticles of $d$-wave superconductors in the vortex lattice
by self-consistently solving the Bogoliubov-de Gennes equations.
It is found for a pure $d_{x^2-y^2}$ state that: (i) low-energy
quasi\-particle bands in the magnetic Brillouin zone
have rather large dispersion even in low magnetic fields,
indicating absense of bound states for an isolated vortex; (ii)
in finite fields with $k_{\rm F}\xi_0$ small,
the calculated tunneling conductance at the vortex core shows
a double-peak structure near zero bias,
as qualitatively consistent with the STM experiment by Maggio-Aprile
{\em et al}.\ $[$Phys.\ Rev.\ Lett.\ {\bf 75} (1995) 2754$]$.
We also find that mixing of a $d_{xy}$- or an $s$-wave component,
if any, develops gradually without transitions
as the field is increased,
having little effect on the tunneling spectra.}
}
\end{center}

\vspace{8mm}
]

Quasiparticles in superconductors have been a matter of a central
theoretical interest in condensed matter physics, underlying almost all
phenomena of superconductors. With a growing number of experiments in zero
magnetic field which support the $d_{x^2-y^2}$-wave scenario for the high-$%
T_c$ cuprates\cite{Shen,Hardy,Wollman,Moler}, efforts have also been made to
clarify their properties in finite magnetic fields. Experimentally, the
scanning tunneling microscope (STM) is one of the most powerful tools to
study them. It was Hess {\em et al}.\ who clarified its potential to probe
the quasiparticles\cite{Hess}. Choosing a conventional $s$-wave material NbSe%
$_2$, they obtained a beautiful vortex-core spectrum with a characteristic
zero-bias peak. Its origin was soon attributed, through numerical
calculations based on the Bogoliubov-de Gennes (BdG) equations for an
isolated vortex\cite{Shore,Gygi}, to the bound quasiparticles around the
vortex core predicted long ago by Caroli {\em et al}.\cite{Caroli}. Although
a direct observation of the discrete levels is yet to be performed, the
experiment confirmed the existence of bound states in the $s$-wave pairing
and stimulated further theoretical studies on quasiparticles in the vortex
states as well as similar experiments on high-$T_c$ materials.

The first microscopic theoretical study on a $d$-wave pairing 
was carried out by Wang
and MacDonald using a lattice model in finite magnetic fields\cite{Wang}.
With parameters chosen appropriate for the cuprates, the calculated
differential tunneling conductance 
at the vortex core shows a single broad peak near zero bias,
whereas that of the corresponding $s$-wave model exhibits a clearly resolved
bound-state structure. On the other hand, the measurement on YBa$_2$Cu$_3$O$%
_{7-\delta}$ by Maggio-Aprile {\em et al}.\ reveals a
double-peak structure around zero bias, which they attributed by using the $s
$-wave result to a couple of bound states split widely due to the large
energy gap\cite{Maggio}. 
This discrepancy between the theory and the experiment raised a
couple of questions: (i) whether or not bound states exist for the $%
d_{x^2-y^2}$ pairing where the gap closes along lines; (ii) why the
double-peak structure appears. Himeda {\em et al}.\ focused on the
strong-correlation
effect to answer the second question\cite{Himeda}.
Choosing the two-dimensional $t$-$J$ model, using a Gutzwiller wavefunction,
and adopting the Gutzwiller approximation, they reported that a double-peak
structure may show up in the low-doping region due to the induced $s$-wave
component which brings a fully opened gap in the core region. However, the
validity of using the Gutzwiller approximation is not well understood here.
On the other hand, Franz and Te\v sanovi\'c investigated an isolated vortex
of a continuum model through a direct diagonalization of a large matrix
\cite{Franz}. 
They have shown that: (i) no bound states exist in the $d_{x^2-y^2}$
model; (iia) the pure $d_{x^2-y^2}$
model also yields a single broad peak near zero bias as
inconsistent with the experiment; (iib) the observed 
double-peak structure may be
explained by a $d_{x^2-y^2}\!+ id_{xy}$ model where the gap opens for an
arbitrary direction so that there exist bound states in the core region.
However, the model (iib) suffers from an inconsistency with the experiments 
that the gap is open everywhere even in zero magnetic field.

With these backgrounds, this letter focuses on a more detailed study on a
continuum $d$-wave model. Due to technical difficulties, investigations on
this fundamental model started rather recently, and only the special case of
an isolated vortex has been considered by now\cite{Franz,Morita,Franz97}. We
here consider the whole range $H_{c1}\!\leq\! H\!\leq\!H_{c2}$ in a unified
way towards a thorough understanding of the basic $d$-wave model, which will
also be useful when treating lattice models with strong correlations or
disorders in the vortex lattice. The technique we use may be called
``Landau-level expansion method,'' developed firstly for the $s$-wave
vortex lattice of $\kappa\!\gg\! 1$\cite{Rajagopal92,NAM92,NAM95}, and then
suitably generalized to include anisotropic pairings and the
magnetic-field variation\cite{Kita}. We thereby clarify those effects caused
by the formation of the vortex lattice and seek alternative mechanisms for
the double-peak structure. We also consider the possibility of mixing of a $%
d_{xy}$- and an $s$-wave component into the pure $d_{x^2-y^2}$ state to see
how it develops with the field strength and how it affects the physical
properties. This topic has attracted much attention recently due to an
observation of a plateau in thermal conductivity $\kappa(H)$\cite{Kris},
followed by a theoretical proposal that it may imply a phase transition
between the $d_{x^2-y^2}$ and the $d_{x^2-y^2}\!+ id_{xy}$ states\cite
{Laughlin}. The origin of the plateau is still controversial\cite{Aubin},
and it will be worth clarifying theoretically whether or not the transition
is really possible.

Our starting point towards those purposes 
is the BdG equations for the eigenfunctions 
${\bf u}_{s}$ and ${\bf v}^{*}_{s}$ labeled by the quantum number $s$ with a
positive eigenvalue $E_s$: 
\begin{eqnarray}
&& \int \! d{\bf r}^{\prime}\left[ 
\begin{array}{cc}
\vspace{1mm}\,\,\, \underline{{\cal H}}({\bf r},{\bf r}^{\prime}) & \,\,\, 
\underline{\Delta} ({\bf r},{\bf r}^{\prime}) \\ 
-\underline{\Delta}^{\! *}\!({\bf r},{\bf r}^{\prime}) & -\underline{{\cal H}%
}^{*}\!({\bf r},{\bf r}^{\prime})
\end{array}
\right] \left[ 
\begin{array}{c}
\vspace{1mm} \,\,{\bf u}_{s}({\bf r}^{\prime}) \\ 
-{\bf v}^{*}_{s}({\bf r}^{\prime})
\end{array}
\right] \hspace{5mm}  \nonumber \\
&& \hspace{25mm}= E_{s}\! \left[ 
\begin{array}{c}
\vspace{1mm} \,\,{\bf u}_{s}({\bf r}) \\ 
-{\bf v}^{*}_{s}({\bf r} )
\end{array}
\right] \, .  \label{BdG}
\end{eqnarray}
Here $\underline{\Delta}$ is the pair potential and $\underline{{\cal H}}$
denotes the normal-state Hamiltonian giving a quadratic dispersion in zero
magnetic field\cite{comm1}; both are $2\!\times\!2$ matrices to include the
spin degrees of freedom. The pair potential is determined self-consistently
by 
\begin{eqnarray}
\underline{\Delta}({\bf r},{\bf r}^{\prime}) = V({\bf r}\! -\! {\bf r}%
^{\prime}) \, \underline{\Phi}({\bf r},{\bf r}^{\prime}) \, ,  \label{Ddef}
\end{eqnarray}
where $V$ denotes the interaction and $\underline{\Phi}$ is the order
parameter defined by 
\begin{eqnarray}
\underline{\Phi}({\bf r},{\bf r}^{\prime}) \equiv\! \sum_s \frac{{\bf u}_s(%
{\bf r}){\bf v}_s^{{\rm T}}({\bf r}^{\prime}) \!-\!{\bf v}_s({\bf r}){\bf u}%
_s^{{\rm T}}({\bf r}^{\prime})}{2} \tanh\!\frac{E_s}{2k_{{\rm B}}T} \, ,
\end{eqnarray}
with $^{{\rm T}}$ denoting the transpose.

To solve these equations efficiently for the vortex-lattice states, we use
the Landau-level expansion method mentioned above, which is sketched as follows
\cite{Kita}. Consider specifically that the field is along the $c$ axis of a
layered two dimensional material. We fix the mean flux density $B$ rather
than the external field $H$, and expand ${\bf u}_{s}$ and ${\bf v}_{s}$ in
the eigenfunctions of the magnetic translation group whose unit cell covers
the area $\phi _{0}/B$ ($\phi _{0}$: the flux quantum). Those eigenfunctions 
$\{\psi _{N{\bf k}\alpha }({\bf r})\}$ are labeled by the Landau-level index 
$N$, the magnetic Bloch vector ${\bf k}$, and the quantum number $\alpha $ $%
(=\!1,2)$ which signifies two-fold degeneracy of the orbital states. Now,
Eq.\ (\ref{BdG}) can be solved separately for each $({\bf k},\alpha )$ due
to the translational symmetry of the vortex lattice; the corresponding
eigenstate is labeled explicitly by $s\!=\!(\nu {\bf k}\alpha \sigma )$ with 
$\nu $ ($\sigma $) the band (spin) index. On the other hand, Eq.\ (\ref{Ddef}%
) is handled via a double expansion of $\underline{\Delta }$ and $%
\underline{\Phi }$ with respect to the center-of-mass and the relative
coordinates. Convenient bases to this end are given by $\{\psi _{N{\bf q}%
}^{({\rm c})}(\frac{{\bf r}+{\bf r}^{\prime }}{2})\}$ and $\{\psi _{Nm}^{(%
{\rm r})}({\bf r}\!-\!{\bf r}^{\prime })\}$, respectively, with ${\bf q}$
another magnetic Bloch vector and $m$ the angular momentum along the $c$
axis $(m\!=\!-N,-N\!+\!1,-N\!+\!2,\cdots )$. The overlap between $\psi _{N_{%
{\rm c}}{\bf q}}^{({\rm c})}(\frac{{\bf r}+{\bf r}^{\prime }}{2})\psi _{N_{%
{\rm r}}m}^{({\rm r})}({\bf r}\!-\!{\bf r}^{\prime })$ and $\psi _{N{\bf k}%
\alpha }({\bf r})\psi _{N^{\prime }{\bf k}^{\prime }\alpha ^{\prime }}({\bf r%
}^{\prime })$ is calculated rather easily, vanishing unless ${\bf q}\!=\!%
{\bf k}+{\bf k}^{\prime }$ and $\alpha _{1}\!=\!\alpha _{2}$. Also worth
noting here is that\cite{Kita2}: (i) a single ${\bf q}$ suffices to describe
the conventional vortex lattice due to its broken translational symmetry;
(ii) $N_{{\rm c}}$'s for the conventional hexagonal (square) lattice are
multiples of $6$ ($4$). Finally $\langle {Nm}|V|{%
N^{\prime }m^{\prime }}\rangle $ is estimated most conveniently in the
momentum representation. To this end, we expand the pair scattering
amplitude in zero field: $V_{{\bf p}{\bf p}^{\prime }}\!\equiv \!(2\pi
)^{2}\langle {\bf p}|V|{\bf p}^{\prime }\rangle $ as $V_{{\bf p}{\bf p}%
^{\prime }}\!=\!\sum_{mm^{\prime }}V^{(m,m^{\prime })}(p,p^{\prime })\,{\rm e}^{
im\varphi _{{\bf p}}\!-\!im^{\prime }\varphi _{{\bf p}^{\prime }}}$, where $%
\varphi _{{\bf p}}$ denotes the angle of the wavevector ${\bf p}$ with the $a
$ axis. Then the matrix element in finite fields 
is obtained as $\langle {Nm}|V|{N^{\prime
}m^{\prime }}\rangle \!=i^{m-m^{\prime }}(-1)^{N+N^{\prime }}V^{(m,m^{\prime
})}(p,p^{\prime })/{4\pi l_{{\rm B}}^{2}}$ with $l_{{\rm B}}$ the
magnetic length and $p\!=\!\sqrt{N}/l_{B}$. Once $V_{{\bf p}{\bf p}^{\prime
}}$ is given explicitly, we can thereby calculate the properties of the
vortex-lattice state over $H_{c1}\!\leq \!H\!\leq \!H_{c2}$.

The pairing interaction we use is given by 
\begin{eqnarray}
N_{0}V_{{\bf p}{\bf p}^{\prime }} &=&-2\left[ \,(p/k_{{\rm F}})^{2}(p^{\prime
}/k_{{\rm F}})^{2}(\,g_{x^{2}-y^{2}}\cos 2\varphi _{{\bf p}}\cos 2\varphi _{%
{\bf p}^{\prime }}\right.   \nonumber \\
&&\left. +g_{xy}\sin 2\varphi _{{\bf p}}\sin 2\varphi _{{\bf p}^{\prime
}})+g_{s}\,\right] \Theta (p)\Theta (p^{\prime })\,,  \label{PairInt}
\end{eqnarray}
where $N_{0}$ is the density of states for both spin in zero field, 
$g$'s are the coupling
constants, $k_{{\rm F}}$ is the Fermi wavenumber, and $\Theta (p)$ denotes a
smooth cut-off\cite{NAM95}: $\Theta (p)\!=\!{\rm e}^{-[(p^{2}-k_{{\rm F}%
}^{2})/p_{c}^{2}]^{4}}$ with $p_{c}$ a cut-off wavenumber. We then utilize
the quantity $T_{c0}^{(j)}\!=\!T_{c0}^{(j)}(g_{j},p_{c})$ ($%
j\!=\!x^{2}\!-\!y^{2}$, $xy$, $s$), defined as the transition temperature in
zero field for the state $j$ in the absence of other two channels, as
parameters to measure the strength of the pairing interaction for the
channel $j$. This procedure may be justified for the weak-coupling model.
Indeed, we find little difference in the calculated results among several
choices of $(g_{j},p_{c})$. Another important parameter in the model is the
coherence length $\xi _{0}\!\equiv \!v_{{\rm F}}/\Delta _{0}$, where $v_{%
{\rm F}}$ is the Fermi velocity and $\Delta _{0}$ denotes the maximum of the
energy gap at $H\!=\!T\!=\!0$. We will clarify below the $k_{{\rm F}}\xi _{0}
$ dependence of the differential tunneling conductance which is
proportional to 
\begin{eqnarray}
N(E,{\bf r}) = - \sum_{s}\, [\,&& |{\bf u}_{s}({\bf r})|^{2}f^{\prime
}(E\!-\!E_{s})
\nonumber \\
+&& |{\bf v}_{s}({\bf r})|^{2}f^{\prime }(E\!+\!E_{s})]\,,
\label{LDOS}
\end{eqnarray}
with $f(E)\!=\!({\rm e}^{E/k_{{\rm B}}T}+1)^{-1}$. Although they can be
included easily, we here neglect: (a) the spatial variation of the magnetic
field since $\kappa $ is quite large for the high-$T_{c}$ materials; (b) the
Pauli paramagnetism for simplicity. The square vortex lattice is assumed
throughout\cite{Won}, and the expansion over $N_{{\rm c}}$ (the
center-of-mass Landau-level index) is cut at some $N_{{\rm c}}^{{\rm max}}$,
typically $N_{{\rm c}}^{{\rm max}}\! = \! 8 \!\sim \! 16$,
with the convergence checked by changing $N_{{\rm c}}^{{\rm max}}$.

We now present the results on the pure $d_{x^2-y^2}$ model ($%
g_s\!=\!g_{xy}\! = \! 0$). Figure 1(a) shows the quasiparticle spectrum in
the magnetic Brillouin zone with $k_{{\rm F}}\xi_0\!=\! 5$ for a
low-temperature and low-field state: $T/T_{c}\!=\! B/H_{c2}\!=\!0.1$. For
comparison, the corresponding one of the $s$-wave model (hexagonal lattice)
is presented in Fig.\ 1(b). We can clearly observe the following difference
between the two. In Fig.\ 1(b) ($s$-wave), a qualitative change in the
dispersion curves is seen around $E_s\!\sim \! \Delta_0$: the curves for $%
E_s\!\gtrsim\! \Delta_0$ are closely spaced with a little dispersion,
whereas the lower-energy bands are flat and occur in pairs with the level
spacing of the order of $\Delta_0^2/\varepsilon_{{\rm F}}$ ($\varepsilon_{%
{\rm F}}$: Fermi energy). As already pointed out by Norman {\em et al}.\cite
{NAM95}, these corresponds to the bound core states of an isolated vortex
with little tunneling probability between adjacent 
\begin{figure}[t]
\begin{center}
\leavevmode
\epsfxsize=84mm
\epsfbox{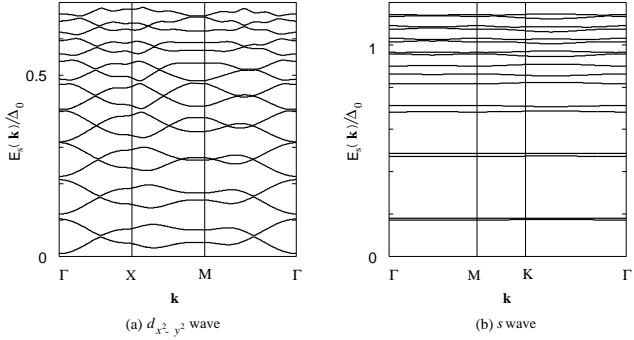}
\end{center}
\caption{Quasiparticle dispersion in the magnetic Brillouin zone for (a) the 
$d_{x^2-y^2}$ wave (square lattice)
and (b) the $s$ wave (hexagonal lattice). Parameters are $T/T_{c}\!=\!0.1$%
, $B/H_{c2}\!=\!0.1$, and $k_{{\rm F}}\protect\xi_0\!=\! 5$ }
\label{fig:1}
\end{figure}
\noindent
cores; one can check that the
number of states for the lowest pair of bands is exactly the same as the
number of vortices. In contrast, the low-energy $d$-wave bands in Fig.\ 1(a)
are densely packed with large dispersion, indicating extended nature of
the corresponding wavefunctions
(this feature
also shows up in the case of the hexagonal lattice).
From this comparison, we may conclude that
no bound states exist for the $d$-wave model even in the zero-field limit of
an isolated vortex, in agreement with the result of Franz and Te\v s%
anovi\'c\cite{Franz}. A remark is necessary at the same time that, strictly
speaking, all the quasiparticles in both models become extended in finite
fields, as long as the vortices are regularly spaced. Thus the difference of
the low-lying states between the two models {\em in finite fields} is of
quantitative (not qualitative) nature.

We turn our attention to the differential tunneling conductance of the pure
$d$-wave model 
and study both $B/H_{c2}$ and $k_{{\rm F}}\xi_0$ dependence of
its spectral line shape. Figure 2 shows the quantity at the vortex core
calculated for $T/T_{c}\!=\! 0.1$ and $k_{{\rm F}}\xi_0\!=\! 5$. 
At the lower field of $B/H_{c2}\!=\! 0.05$, we see a
single broad peak around zero bias as
previous calculations\cite{Wang,Franz}.
It indicates the existence of virtually bound quasiparticles in the core
region. As the field is increased, however, the line shape starts changing
to show a clear double-peak structure at $B/H_{c2}\!=\! 0.3$, which
is qualitatively consistent with the spectrum observed by Maggio-Aprile {\em %
et al}.\cite{Maggio}. Though not presented here, such a change cannot be
seen in the spectra obtained for $k_{{\rm F}}\xi_0\!=\! 10$ where the single
broad peak remains essentially unchanged over a wide range of the field
strength. Hence the splitting may be considered as inherent to systems with
small values of $k_{{\rm F}}\xi_0$. It is explained here in connection with
the increase of the quasiparticle hopping probability between cores as
follows. The equilibrium thermodynamic potential at $T\!=\! 0$ can be
written in terms of Eq.\ (\ref{LDOS}) as
\begin{eqnarray}
\Omega = \int \! d{\bf r} \! \int^0_{-\infty} \! d\omega \, \omega \, N({\bf %
r},\omega) + {\rm Tr}\, \underline{\Phi}^{\ast} \underline{\Delta} \, ,
\end{eqnarray}
\begin{figure}[tbp]
\begin{center}
\leavevmode
\epsfxsize=70mm
\epsfbox{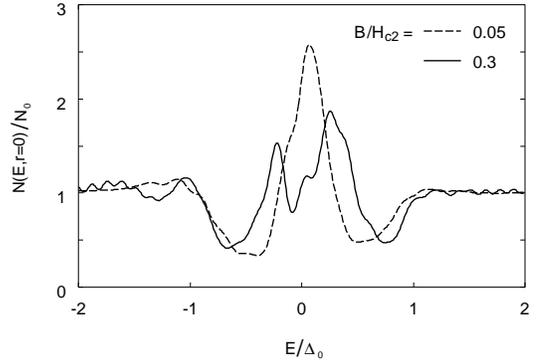}
\end{center}
\caption{The
tunneling conductance of the $d_{x^2-y^2}$-wave model 
at the core center for $B/H_{c2}\!=\!0.05$ and $0.3$
with $T/T_c\!=\!0.1$ and $k_{{\rm F}}\protect\xi_0\!=\! 5$.}
\label{fig:2}
\end{figure}
\noindent
where Tr implies integration and summation over the space 
and the spin coordinates, respectively. The second 
term on the right-hand side is expected
not to depend strongly on the details of the quasiparticle spectrum. The first
term, on the other hand, is quite sensitive to the structure of the local
density of states. In particular, high density of states at $\omega\!=\!0$
is energetically unfavorable so that the system tends to reduce it by using
any perturbation available. We identify this perturbation with the hopping
probability of the low-energy quasiparticles, whose magnitude may be
measured by the dimensionless distance between adjacent vortices, $k_{{\rm F}%
}l_{B}\!\sim\! k_{{\rm F}}\xi_{0}/\sqrt{B/H_{c2}}$.
Hence the identification
tells us that the double-peak structure becomes clearer as $k_{{\rm F}%
}\xi_{0}$ ($B/H_{c2}$) decreases (increases), which is consistent with our
numerical results. Our calculations performed for other values of $k_{{\rm F}%
}\xi_{0}$ reveal that the peak splitting shows up at intermediate fields for 
$k_{{\rm F}}\xi_{0}\! \lesssim\! 5$. Since $k_{{\rm F}}\xi_{0}$ is of the
order of unity for the high-$T_c$ materials, the above mechanism may well
explain the tunneling spectrum observed at $H\!=\! 6$T\cite{Maggio}. A
detailed experiment on the field dependence may be
worth carrying out.

We finally consider the mixing of a $d_{xy}$-wave or an $s$-wave component
into the dominant $d_{x^{2}-y^{2}}$ wave to study how it affects
physical properties in finite magnetic fields. Such a state has been
suggested to be realized in finite fields at low temperatures accompanying a
phase transition\cite{Kris,Laughlin}. Franz and Te\v{s}anovi\'{c} treated
this problem in the zero-field limit 
of an isolated vortex, where parameters were
chosen in such a way that $d_{x^{2}-y^{2}}\!+id_{xy}$ state is already
formed in zero magnetic field. Clearly this is not consistent with various
experiments\cite{Shen,Hardy,Wollman,Moler} which indicate a $d_{x^{2}-y^{2}}$
pairing in zero field. We hence choose the parameter $g_{xy}$ or $g_{s}$ in
Eq.\ (\ref{PairInt}) in such a way that a pure $d_{x^{2}-y^{2}}$
state is stabilized
in zero field. More specifically, $T_{c0}^{(xy)}/T_{c}$ 
($T_{c0}^{(s)}/T_{c}$) is put $0.31$ ($0.36$) in the
following $d_{xy}$-wave ($s$-wave) study,
whereas the critical value is 0.33 (0.38) 
above which $d_{x^{2}-y^{2}}\!+id_{xy}$ ($d_{x^{2}-y^{2}}\!+is$) 
state is formed in zero field below a certain critical temperature
$T_{c}^{(xy)}$ ($T_{c}^{(s)}$) lower than $T_{c0}^{(xy)}$ 
($T_{c0}^{(s)}$).
Notice incidentally that $T_{c}\!=\! T_{c0}^{(x^2-y^2)}$.\hfill
Figure 3 shows the field dependence
\begin{figure}[tbp]
\begin{center}
\leavevmode
\epsfxsize=70mm
\epsfbox{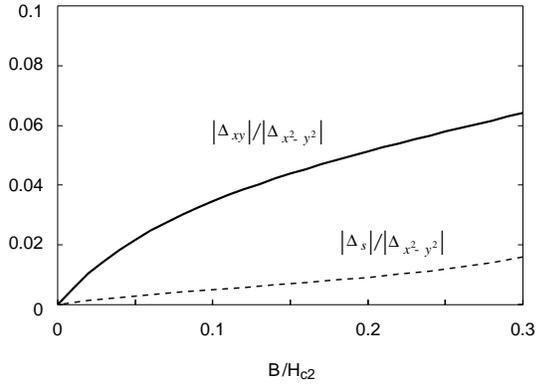}
\end{center}
\caption{Field dependence of the ratio $|\Delta_{xy}|/|\Delta_{x^2-y^2}|$ 
($|\Delta_{s}|/|\Delta_{x^2-y^2}|$)
at the center of the vortex-core square
with $T/T_c\!=\!0.1$, $k_{{\rm F}}\protect\xi_0\!=\! 5$, and $%
T_{c0}^{(xy)}/T_{c}\!=\! 0.31$ ($T_{c0}^{(s)}/T_{c}\!=\! 0.36$).}
\label{fig:3}
\end{figure}

\noindent
of the maximum-gap ratios
$|\Delta_{xy}|/|\Delta _{x^{2}-y^{2}}|$ and
$|\Delta_{s}|$ $/|\Delta_{x^{2}-y^{2}}|$ at the midpoint of the
next-nearest-neighbor vortices of the square lattice.
One can see clearly that the component $|\Delta _{xy}|$ develops
gradually without any transitions with the field strength. This fact
can also be realized on a purely group-theoretical ground as follows. There
are four mirror reflections in zero field which distinguish the $d_{xy}$
state from the $d_{x^{2}-y^{2}}$ state. Those symmetry operations, however,
no longer exist at finite fields due to the supercurrent flowing in the
system. It hence follows that the mixing starts immediately above $H_{c1}$.
The same is true for the $s$-wave mixing as already pointed out\cite
{Ren,Berlinsky}. The result implies that the reason for the observed plateau
in thermal conductivity\cite{Kris,Aubin}, if it does exist conclusively,
should be sought elsewhere. Finally, Fig.\ 4 compares the tunneling
conductance of the $d_{x^{2}-y^{2}}\!+id_{xy}$ state (solid line) with that
of the pure $d_{x^{2}-y^{2}}$ state (dashed line) at
$T/T_c\!=\!0.1$. We see little difference
between the two even for $B/H_{c2}\!=\!0.3$ and $k_{{\rm F}}\xi _{0}\!=\!5$.
A similar result has been found for the case of the $s$-wave component.
These results suggest that the mixing without a transition brings only small
changes for the physical properties in finite magnetic fields.

In summary, we have carried out fully self-consistent calculations
of the BdG equations for a basic $d$-wave model with continuously
varying the field strength. 
We have thereby presented several new results
connected closely with the detailed nature of the $d$-wave quasiparticles
in the vortex lattice.
The method used here will be helpful for the microscopic 
understanding of
other anisotropic superconductors in finite magnetic
fields as well as rotating Fermi superfluids.

One of the author (T.K.) is indebted to Ch.\ Renner for a useful discussion.
Numerical calculations were performed on an Origin 2000 in "Hierarchical
matter analyzing system" at the Division of Physics, Graduate School of
Science, Hokkaido University.

\begin{figure}[tbp]
\begin{center}
\leavevmode
\epsfxsize=70mm
\epsfbox{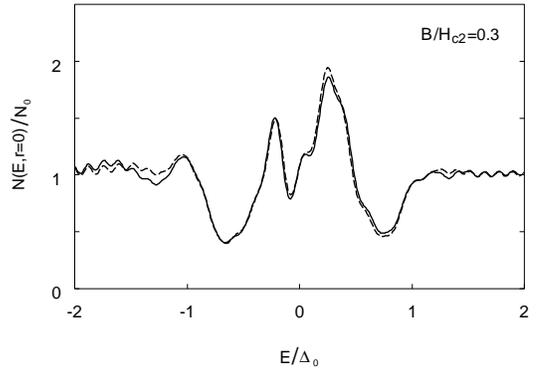}
\end{center}
\caption{The tunneling conductance at the core center for the 
$d_{x^2-y^2}\!+id_{xy}$ state (solid line) compared with that 
of the pure $d_{x^2-y^2}$ state (dashed line).
Parameters are $B/H_{c2}\!=\!0.3$, $T/T_c\!=\!0.1$,
$k_{{\rm F}}\xi _{0}\!=\!5$,
and $T_{c0}^{(xy)}/T_{c}%
\!=\! 0.31$.}
\label{fig:4}
\end{figure}



\begin{references}
\bibitem{Shen}  Z.-X. Shen {\em et al}., Phys. Rev. Lett. {\bf 70} 1553,
(1993).

\bibitem{Hardy}  W.N. Hardy {\em et al}., Phys. Rev. Lett. {\bf 70},
3999  (1993).

\bibitem{Wollman}  D.A. Wollman {\em et al}., Phys. Rev. Lett. {\bf 71},
2134 (1993).

\bibitem{Moler}  K.A. Moler {\em et al}., Phys. Rev. Lett. {\bf 73}, 2744
(1994).

\bibitem{Hess}  H.F. Hess {\em et al}., Phys. Rev. Lett. {\bf 62},  214
(1989).

\bibitem{Shore}  J.D. Shore {\em at al}., Phys. Rev. Lett. {\bf 62},  3089
(1989).

\bibitem{Gygi}  F. Gygi and M. Schl\H uter, Phys. Rev. B. {\bf 43}, 
7609 (1991).

\bibitem{Caroli}  C. Caroli, P.G. de Gennes, and J. Matricon,  
Phys. Lett. {\bf 9}, 307
(1964).

\bibitem{Wang}  Y. Wang and A.H. MacDonald, Phys. Rev. B {\bf 52}, R3876
(1995).

\bibitem{Maggio}  I. Maggio-Aprile {\em et al}., Phys. Rev. Lett. {\bf 75%
}, 2754 (1995).

\bibitem{Himeda}  A. Himeda {\em at al}., J. Phys. Soc. Jpn. {\bf 66},
3367 (1997).

\bibitem{Franz}  M. Franz and Z. Te\v sanovi\'c, Phys. Rev. Lett. {\bf %
80}, 4763 (1998).

\bibitem{Morita}  Y. Morita, M. Kohmoto, and K. Maki, 
Phys. Rev. Lett. {\bf 78},
4841 (1997).

\bibitem{Franz97}  M. Franz and M. Ichioka, Phys. Rev. Lett. {\bf 79},
4513 (1997).

\bibitem{Rajagopal92}  A.K. Rajagopal, Phys. Rev. B{\bf 46}, 1224 (1992).

\bibitem{NAM92}  M. R. Norman, H. Akera,  and A.H. MacDonald,
Physica C{\bf 196}, 43 (1992).

\bibitem{NAM95}  M. R. Norman, A.H. MacDonald and H. Akera,
Phys. Rev. B {\bf 51} (1995) 5927.

\bibitem{Kita}  T. Kita, J. Phys. Soc. Jpn. {\bf 67}, 2075 (1998).

\bibitem{Kris}  K. Krishana {\em et al}., Science {\bf 277}, 83 (1997).

\bibitem{Laughlin}  R.B. Laughlin, Phys. Rev. Lett. {\bf 80}, 5188
(1998).

\bibitem{Aubin}  H. Aubin {\em et al}., Phys. Rev. Lett. {\bf 82}, 624
(1999).

\bibitem{comm1}  This restriction may be removed by using the
Onsager-Lifshitz semiclassical quantization scheme.

\bibitem{Kita2}  T. Kita: J. Phys. Soc. Jpn. {\bf 67}, 2067 (1998).

\bibitem{Won}  H. Won and K. Maki, Phys. Rev. B {\bf 53}, 5927 (1996).

\bibitem{Ren}  Y. Ren, J.-H. Xu, and C.S. Ting, 
Phys. Rev. Lett. {\bf 74}, 3680
(1995).

\bibitem{Berlinsky}  A.J. Berlinsky {\em et al}., Phys. Rev. Lett. {\bf %
75}, 2200 (1995).
\end{references}
\end{document}